\documentclass[12pt,a4]{article}

\usepackage{amsmath}
\usepackage{amssymb}
\usepackage{graphicx}
\usepackage{color}

\newcommand{\rf}[1]{(\ref{#1})}
\newcommand{\beq}{\begin{equation}}
\newcommand{\eeq}{\end{equation}}
\newcommand{\bea}{\begin{eqnarray}}
\newcommand{\eea}{\end{eqnarray}}

\newcommand{\e}{{\rm e}}

\newcommand{\equ}{\!=\!}
\newcommand{\plu}{\!+\!}

\newcommand{\g}{\gamma}

\newcommand{\lam}{\lambda}
\newcommand{\Lam}{\Lambda}
\renewcommand{\b}{\beta}
\newcommand{\al}{\alpha}

\newcommand{\ep}{\varepsilon}

\newcommand{\Om}{\Omega}
\newcommand{\del}{\delta}
\newcommand{\Del}{\Delta}
\newcommand{\sg}{\sigma}
\newcommand{\kp}{\kappa}

\newcommand{\oh}{\frac{1}{2}}

\newcommand{\dg}{\dagger}

\newcommand{\ra}{\rangle}

\newcommand{\prt}{\partial}

\newcommand{\cH}{{\cal H}}

\def\void{}
\def\labelmark{}

\newenvironment{formula}[1]{\def\labelname{#1}
\ifx\void\labelname\def\junk{\begin{displaymath}}
\else\def\junk{\begin{equation}\label{\labelname}}\fi\junk}%
{\ifx\void\labelname\def\junk{\end{displaymath}}
\else\def\junk{\end{equation}}\fi\junk\labelmark\def\labelname{}}

{\ifx\void\labelname\def\junk{\end{array}\end{displaymath}}
\else\def\junk{\end{array}\right.\end{equation}}
\fi\junk\labelmark\def\labelname{}\def\junk{}
}

\newcommand{\beqv}{\begin{formula}{}}

\vspace{12pt}

\newcommand{\hH}{{\hat{H}}}

\newcommand{\fm}{f_{\rm mod}}
\newcommand{\fg}{f_{\rm gcdt}}
\newcommand{\fds}{f_{\rm ds}}

\begin{document}

\rightline{\today}

\begin{center}
\vspace{24pt}
{ \Large \bf Is the present acceleration of the Universe caused by 
  merging with other universes?}

\vspace{24pt}

{\sl J.\ Ambj\o rn}$\,^{a,b}$,
and {\sl Y.\ Watabiki}$\,^{c}$

\vspace{10pt}

{\small

$^a$~The Niels Bohr Institute, Copenhagen University\\
Blegdamsvej 17, DK-2100 Copenhagen \O , Denmark.\\
email: ambjorn@nbi.dk
\vspace{10pt}

$^b$~Institute for Mathematics, Astrophysics and Particle Physics
(IMAPP)\\ Radbaud University Nijmegen, Heyendaalseweg 135, 6525 AJ, \\
Nijmegen, The Netherlands

\vspace{10pt}

$^c$~Tokyo Institute of Technology,\\ 
Dept. of Physics, High Energy Theory Group,\\ 
2-12-1 Oh-okayama, Meguro-ku, Tokyo 152-8551, Japan\\
{email: watabiki@th.phys.titech.ac.jp}

}

\end{center}

\vspace{24pt}

\begin{center}
{\bf Abstract}
\end{center}

\noindent
We show that by allowing our Universe to merge with other universes  
one is lead to  modified Friedmann equations
that  explain the present accelerated expansion of our Universe without the need of a cosmological constant.

\newpage

\section{Introduction}\label{intro}

We have proposed a modified Friedmann equation \cite{aw2,aw3} 
which changes the late time cosmology, such that one 
does not need a cosmological constant to explain the present day acceleration of our Universe. While it is our hope that
this modified Friedmann  equation can eventually be derived from an 
underlying microscopic theory \cite{aw1}, we will here treat it 
as a phenomenological model that we can obtain in  a simple way from the standard Friedmann equation\footnote{It should be noted that ideas somewhat related  to the ones 
presented here and in \cite{aw2,aw3,aw1}  have also been advocated in \cite{newkawai}}.

 Our starting point is the Hartle-Hawking minisuperspace action, 
 which after the rotation of the conformal factor can be written as 
 \beq\label{j1}
 S=  \int dt  \, \Big( \frac{\b\dot{v}^2}{2N v} + \lam N v\Big).
 \eeq
 Before rotating to  Euclidean spacetime 
 and the rotation of the conformal factor, this action 
 is just the standard minisuperspace action
 \beq\label{j2}
 S =  \int dt  \,\Big(-\frac{\b\dot{v}^2}{2Nv} - \lam N v\Big),
 \eeq
 written using the metric
 \beq\label{j3}
 ds^2 = -N^2(t)dt^2 + a^2(t) d\Om_d,\qquad d \Om_d = \sum_{i=1}^d dx_i^2.
 \eeq
 In \rf{j1} and \rf{j2} we are using units where $c=\hbar = 1$ and  the constant 
 $\b = (d-1)/d$,  where $d$ is the dimension of space. We have used
 \beq\label{j3a} 
 v(t) = \frac{1}{\kp} a^d(t),\quad \kp = 8\pi G, \quad \mbox{$G=$ the gravitational constant.}
 \eeq
  as variable rather than the scale factor $a(t)$. $v(t)$ is proportional 
 to the spatial $d$-volume at time $t$ and below we will just call it the $d$-volume.
 For the  Hubble parameter $H(t)$ we then have
 \beq\label{j3b}
 H(t) \equiv \frac{\dot{a}}{a} = \frac{1}{d} \, \frac{\dot{v}}{v}.
 \eeq
 The reason we prefer to use $v(t)$ instead of $a(t)$ is that the minisuperspace 
 action \rf{j1}-\rf{j2} is then valid in  all space dimensions $d>1$. 
 Finally $ \lam$ denotes the cosmological constant. Presently we will ignore 
 matter, only assume a cosmological constant term. Below we will include matter 
 in our considerations.
  
  It should be noted that \rf{j1} also appears as the leading term in an effective action 
 in a model of four-dimensional quantum gravity 
 known as Causal Dynamical Triangulations (CDT). In that case one is not 
 assuming a minisuperspace reduction, but performs the path integral over all degrees of freedom except $v(t)$. Thus \rf{j1}
 might be more general than suggested by the minisuperspace reduction.  
 This CDT result is a numerical result, obtained via Monte Carlo Simulations of the 
 CDT lattice gravity model where one  identifies $G/\ep^2$ and $\lam \ep^4$ with 
 the corresponding dimensionless lattice coupling constants, $\ep$ denoting the 
 length of the lattice links, i.e.\ the UV lattice cut-off (see \cite{4dcdt}, and 
 \cite{review} for  reviews). Quite remarkably, for CDT in $d=1$, \rf{j1} can be derived
 analytically  (with $\b \neq 0$), and it is an effective action coming entirely from 
 the path integral measure \cite{al}, since the classical Einstein action is a topological 
 invariant in the case of $d=1$.

The classical Hamiltonian corresponding to \rf{j2} is\footnote{\label{footnote2}
Since there is no  
time derivative of $N(t)$ in \rf{j2}, we should strictly speaking treat it as a constraint 
system and  we can choose as a Hamiltonian $\cH_u = \cH + u(t) P_N$, where $P_N$ is 
the  momentum conjugate to $N$, and impose the constraint $P_N = 0$ in phase
space. This leads to $\cH=0$ on the constraint surface as a consistent secondary constraint
and $\dot{N} = u(t)$, expressing the invariance with respect to  time reparametrization. In the 
following we will use this invariance to choose $N$ constant ($N=1$), 
and then consider ``on shell''
solutions $\cH=0$.  }
\beq\label{j4}
\cH(v,p)  = N v \Big(- \frac{ p^2}{2 \b} + \lam\Big),
\eeq
where $p$ denotes the momentum conjugate to $v$. In the following we will 
be interested in $d=3$,  i.e.\ $2\b=4/3$.

 A classical solution corresponding to the action \rf{j2} is the de Sitter spacetime\footnote{
 \label{footnote-p} Note that the solution of $\cH(v,p) =0$ corresponding to an  expanding universe has 
 $ p = -\sqrt{\lam} < 0$. The reason for the somewhat counter-intuitive negative values of $p$ is the negative  sign of the ``kinetic" terms in \rf{j2} and \rf{j4}.}
 \beq\label{j5}
 v(t) = v(t_0) \, \e^{ \sqrt{ 3 \lam} (t-t_0)}.
 \eeq
 We now want to go beyond this classical picture, but staying as close as possible to the minisuperspace picture, by 
 trying to include the possibility that our Universe can absorb other universes,
 which we denote baby universes even if they are not necessarily small.
 
 \section{Expansion by merging with other universes}

 Since we will allow for other universes to merge with 
 our Universe, we are really discussing a multi-universe theory and like in a many-particle theory it is natural 
 to introduce creation and annihilation operators $\Psi^\dg(v)$ and $\Psi(v)$ 
 for single universes of spatial volume $v$. 
 In a full theory of four-dimensional quantum gravity
the spatial volume alone will of course not provide a complete characterization
of a state at a given time $t$. As mentioned we will here make the drastic simplification
to work in a minisuperspace approximation where the spatial universe {\it is} characterized
entirely by the spatial volume $v$.
 Thus,  denote the quantum state of 
 a spatial universe with volume $v$ by $| v \ra$. We 
 consider now the multi-universe Fock space constructed from such single universe states and denote the Fock vacuum state
 by $| 0 \ra$. Then 
 \beq\label{j11a}
 [\Psi(v),\Psi^\dg(v')] = \del (v-v'), \quad \Psi^\dg(v) | 0\ra = | v\ra, \quad \Psi(v) |0\ra = 0.
 \eeq 
 
 In this way the (minisuperspace) quantum Hamiltonian that includes
 the creation and destruction of universes can be written as \cite{sft}
 \bea
\hH &=& \hH^{(0)} - g \int dv_1 \int dv_2 \;\Psi^\dg(v_1)\Psi^\dg(v_2)\;
(v_1\plu v_2)\Psi(v_1\plu v_2) -
\label{j12} \\ 
&& \!\!\! g\int dv_1 \int dv_2 \;\Psi^\dg(v_1\plu v_2)\;v_2\Psi(v_2)\;v_1\Psi(v_1)
- \!\int \frac{dv}{v}\,  \rho(v) \Psi^\dg (v), \quad \rho(v)= \del(v)\nonumber
\eea
\beq\label{j13}
\hH^{(0)} = \int_0^\infty \frac{dv}{v} \; \Psi^\dg (v) 
\hat{\cH}^{(0)} \, v  \Psi(v),~~~~~\hat{\cH}^{(0)}=  
{v}\Big(-  \frac{3}{4} \frac{d^2}{d v^2}\,+\lam  \Big)
\eeq
${\hH}^{(0)}$ describes the quantum Hamiltonian corresponding to the action \rf{j1} 
(with $\beta = 2/3$) and describes the propagation of a 
single universe, while the two cubic  terms 
describe the splitting of a universe into two and the merging of two universes into one, respectively.  Finally, the last term implies that a universe can be 
created from the Fock vacuum $| 0 \ra$ provided the spatial volume is zero.
If it was not for this term $\hH | 0 \ra =0$, and the Fock vacuum would be stable.
Again, in our minisuperspace approximation we do not attempt to describe
how such a merging or splitting realistically takes place, the only thing which has our interest 
is how the volume of space can be influenced by such merging or splitting processes, and 
for this the minisuperspace model might give us some interesting hints.

\begin{figure}[t]
\centerline{\scalebox{0.18}{\rotatebox{0}{\includegraphics{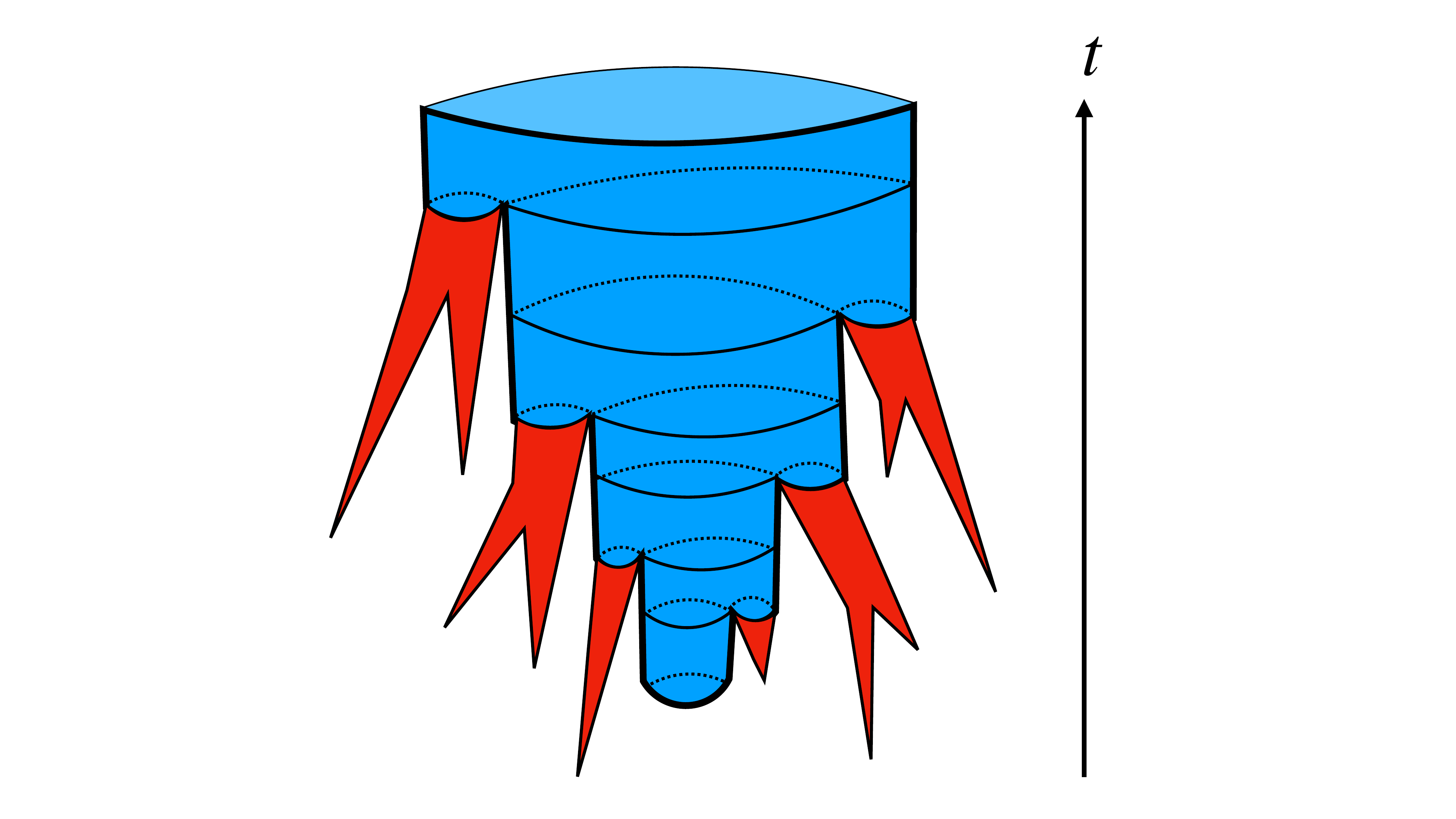}}}}
\caption[figxj1]{{\small
Our Universe (blue), represented as  one-dimensional circles, 
propagating in time, with baby universes (red) merging and increasing the spatial volume.
}}
\label{figxj1}
\end{figure}

Even the minisuperspace Hamiltonian $\hH$ is too complicated to be solved in general.
A universe can successively split in many, be joined by many and  a part 
that splits off can later
rejoin, thereby changing the topology of spacetime. Since the Hamiltonian is essentially
dimension independent (all dimension dependence is absorbed in the coupling constants
$\kp$, $\lam$ and $g$), what we are describing is the so-called 
string field theory of two-dimensional CDT \cite{sft}. For this string field theory there exists
 a  truncation that can be solved analytically \cite{gcdt}\footnote{Surprisingly, some CDT
 string field amplitudes can actually be calculated non-perturbatively to all order in the genus expansion, see \cite{allgenus}.}, called generalized CDT (GDCT),
 and that at the  same time  has our main interest from a cosmological point of view. It 
 follows the evolution of a universe (let us call it ``our'' Universe) in time. During this evolution
 it can merge with other universes (denoted baby universes), 
 created at times $t_i$ with spatial volumes 
 $v_i(t_i) =0$ (see Fig.\ \ref{figxj1}). We will assume these universes have the same 
  coupling constants as our Universe. The effective Hamiltonian, 
 the so-called {\it inclusive} Hamiltonian, first introduced in \cite{inclusive}, governing 
 the evolution of the Universe, is obtained from the path integral by integrating over the times
 $t_i$ and summing over the number of  baby universes merging with our Universe. In the path integral, the various baby universes, 
 characterized at a given time $t$ by spatial volumes $v_i(t)$, can themselves be merged
 with other baby universes. One integrates over all possible $v_i(t)$, not necessarily related
 to solutions of any classical equations.

 Let us describe in some detail how such an 
 inclusive Hamiltonian can be obtained from $\hH$. 
 The simplest way to incorporate 
 an absorption of baby universes into the propagation of our Universe is to replace 
the quantum field $\Psi(v)$, representing the disappearance of a universe of 
spatial volume $v$ when it is absorbed by our Universe, by a classical field $\psi(v)$.
Thus we make the following replacement in the third term on the rhs of eq.  \rf{j12}
\beq\label{j33}
\Psi^\dg(v_1+v_2) \Psi(v_1) \Psi(v_2) \to  
\Psi^\dg(v_1+v_2) \psi(v_1) \Psi(v_2) + \Psi^\dg(v_1+v_2) \Psi(v_1) \psi(v_2)
\eeq
The terms on the rhs of eq.\ \rf{j33} contribute to the propagation of our 
Universe since they are of the form $\Psi^\dg(v_1) \big( \cdot \big) \Psi(v_2)$, but 
contrary to the terms in $\hH_0$ they are non-local in $v$. By Taylor expanding 
$\Psi^\dg(v+w)$ around $v$, this non-locality can be expanded as a power series
in the non-local operator  $(d/dv)^{-1}$. After some algebra this leads to 
\beq\label{j34}
\int dv dw\, \Psi^\dg(v+w) \,w\psi(w) \,v\Psi(v) =
\int dv \Psi^\dg(v)\, F\!\left(\frac{d}{dv}\right) v \Psi(v),
\eeq
where $F(p)$ is the Laplace transform of $\phi(v) = v \psi (v)$:
\beq\label{j35}
 \phi(v) = \phi_0 + \phi_1{v}+\phi_2 {v^2}+ \cdots, \quad
F(p) = \int_0^\infty \!\!\!dv \;\e^{-p \,v} \phi(v) = \frac{\Gamma(1)\phi_0}{ p} + 
\frac{\Gamma(2)\phi_1}{ p^2} + \cdots 
\eeq
Thus $\hat{\cH}_0$ from \rf{j13} will be replaced by the inclusive Hamiltonian 
\beq\label{j36}
\hat{\cH}_{\rm inc} = \hat{\cH}_0 -2g  F\!\left( \frac{d}{dv}\right) {v} 
\eeq
In GCDT $F(p)$ is determined by the self-consistency   requirement that our Universe,
modified by the impact $\phi(v)$ of  baby universes,  should be identical to the 
baby universes it absorbs.  We refer to \cite{gcdt} (or to the Lecture Notes \cite{lectures} ) 
for the details of this determination\footnote{The only difference compared the the 
derivation in \cite{gcdt} is that the sign of $g$ appearing in \rf{j37} and \rf{j38} will be different
from the sign appearing in \cite{gcdt}, the reason being that we consider the 
{\it absorption} rather than the emission of baby universes. In earlier articles
discussing a modified Friedmann equation (\cite{aw2}) we used the notation from 
\cite{gcdt}, but with negative $g$, which effectively meant that we were considering 
absorption rather than emission of baby universes, like here.}. Finally, 
rotating back from the Hartle-Hawking metric to Lorentzian signature, as when going from \rf{j1} to \rf{j2} and replacing\footnote{The rotation from the Hartle-Hawking 
metric to the Lorentzian metric involves a double analytic continuation, 
namely both in time and also a conformal factor rotation, which in the 
minisuperspace metric parametrization becomes $v \to -i v$.} the  $ -id/dv$ by the classical momentum $p$ conjugate to $v$ we obtain the following semiclassical Hamiltonian 
\beq
\cH =  {v}\Big( - \,\frac{3}{4}( p^2 + {\lam} -2 g F( p)\Big)
= \frac{3}{4} {v} \Big( (p +\al)  
\sqrt{ ( p-\al)^2 +\frac{2 g}{\al} }\;\;\Big),
\label{j38}
\eeq
where $\al$ satisfies the equation\footnote{In the case of $g=0$ we choose the solution 
$\al = \sqrt{\lam} >0$.  In this case, as remarked in footnote \ref{footnote-p}, 
$  p= - \al < 0$ and $\sqrt{(p -\al)^2} = \al - p$. Since we consider $g\geq 0$, the solution $\al >0$ will only increase with increasing $g$
and $\cH =0$ leads to $ p = -\al < 0$.}
\beq\label{j7}
\al^3 -  \frac{4}{3} \lam\al - g = 0, 
\eeq
It is seen that if $g=0$ one obtains precisely the $H$ in \rf{j4} (with $N=1$).
A solution to the ``on shell'' Hamiltonian equations is 
\beq\label{j8}
v(t) = v(t_0) \, \e^{ \frac{3}{2}  \Sigma (t-t_0)}, \quad 
\Sigma = \sqrt{ \al^2 + \frac{ g}{2\al}}.
\eeq
Again, if we choose $g \equ 0$ we obtain the de Sitter solution \rf{j5}. 
Increasing $g$ will increase the expansion rate of the 
universe, the intuition behind this being illustrated in Fig.\ \ref{figxj1}, 
and we can actually take the cosmological 
constant $\lam \equ 0$ and still obtain an expanding universe:
\beq\label{j9}
v(t) = v(t_0) \, \e^{ \sqrt{\frac{27}{8}} \, g^{1/3} (t-t_0) }.
\eeq
Thus we have the same classical de Sitter solution as \rf{j5} provided
\beq\label{j10}
g^{2/3}  = \frac{8}{9}\; \lam,
\eeq
but the origin of this exponential expansion is now not a cosmological constant but
instead the ``bombardment'' of our Universe by baby universes. 

\section{Including matter}

Until now we have considered our Universe, but without matter. We will now include matter
in the most simple way, as a matter density $\rho_{\rm m}(v)$ in the 
Hamiltonian \rf{j38}\footnote{The 
inclusion of matter in this way introduces an asymmetry between 
the baby universes and our Universe, in the sense that we do not 
include such a  matter term in the evolution of the baby universes.}. In addition
we will only consider relatively late times in the evolution of 
our universe, namely the period after the time of last scattering ($t_{\rm LS}$), where matter
to a good approximation can be considered as dust that exerts  zero pressure . Thus the 
Hamiltonian has the form
\beq\label{j30}
 \cH [v,p] =  {v} \,( -f( p)+ \kp \rho_{\rm m}(v) \,), 
 \quad v \rho_{\rm m}(v) = v_0 \rho_{\rm m} (v_0), 
\eeq
where $v_0$ and $\rho_{\rm m}(v_0)$ denote the values at the present time $t_0$ and 
where we will later consider three different choices of $f( p)$. Before that, let us 
discuss the solution of the eoms for arbitrary $f( p)$ in \rf{j30}. The 
eoms simplify since  $v \rho_{\rm m}(v)$ is constant.  
\beq\label{j41}
\dot{v} = \frac{\prt \cH}{\prt p} = -v f'( p),\quad {\rm i.e.} \quad  3 \,\frac{\dot{a}}{a} =
\frac{\dot{v}}{v} = -f'( p)
\eeq
\beq\label{j42}
~~\dot{ p} = - \frac{\prt \cH}{\prt v } =  f( p), \quad {\rm i.e.} \quad 
t- t_{\rm LS} = \int_{ p_{\rm LS}}^{ p} \frac{d \,  p}{f( p)}
\eeq
By construction any solution to \rf{j41}-\rf{j42} will satisfy $\cH = {\rm const}$, and we are 
interested in the ``on-shell'' solutions $\cH=0$, which by \rf{j30} implies that 
\beq\label{j40}
f (p) = \kp \rho_{\rm m}(v) = \kp \rho_{\rm m}(v_0) \frac{ v_0}{v} =
f( p_0) \frac{ v_0}{v} = f( p_0)(1+z)^3, 
\eeq
where $p_0$ denotes the value of $p$ at present time $t_0$ and 
$z$ denotes the redshift at time $t$, i.e. 
\beq\label{j40a}
z(t)+1 = \frac{a(t_0)}{a(t)} = \Big(\frac{v(t_0)}{v(t)}\Big)^{1/3}.
\eeq
{\it Eq.\ \rf{j40} is the generalized Friedmann when eq.\ \rf{j41} is used to express 
$ p$ in terms of ${\dot{a}}/{a}$.} 

Using $ p$ as a parameter instead of $t$ (the relation between the two parameters 
 is defined by eq.\ \rf{j42})
we can immediately write parametric expressions for 
a number of relevant functions expressed in terms of $f( p)$. Let us define these.
The redshift $z(t)$ or $z( p)$ is 
\beq\label{def1}
z = \frac{a(t_0)}{a(t)}-1 = \Big(\frac{f( p)}{f( p_0)}\Big)^{\frac{1}{3}}\!\! -1.
\eeq
The Hubble parameter $H(t)$ or $H(p)$ 
is defined as (see also eq.\ \rf{j3b})
\beq\label{def2}
H= \frac{\dot{a}(t)}{a(t)} = - \,\frac{1}{3} \, f'( p).
\eeq
The {\it formal density} $\rho_f(t)$ or $\rho_f( p)$ related to the function $f( p)$ 
is  obtained by writing the generalized Friedmann equation \rf{j40} as 
\beq\label{def3}
\Big(\frac{\dot{a}(t)}{a(t)}\Big)^2 = \frac{\kp \rho_{\rm m}(v)}{3} + \frac{\kp \rho_f(v)}{3}
\eeq
from which we deduce (using the eoms) 
\beq\label{def4}
\kp \rho_f ( p) = \frac{1}{3} \big(f'( p)\big)^2 - f( p).
\eeq
\beq\label{def5}
\kp \frac{d {\rho}_f}{d t} = f( p) f'( p) \Big( \frac{2}{3} f''( p) -1\Big).
\eeq
We define the {\it formal pressure} $P_{ f}$ related to $\rho_{ f}$ 
by the energy conservation equation 
\beq\label{def6}
\frac{d}{dt} ( v \rho_f) +  P_f \frac{d}{dt} v = 0.
\eeq
This leads to 
\beq\label{def7}
P_f = f( p)   \Big( \frac{2}{3} f''( p) -1\Big) - \rho_f(v)
\eeq
and the {\it formal equation of state parameter} $w_f$ is defined (for $\rho_f \neq 0$) by
\beq\label{def8}
w_f  = \frac{P_f}{\rho_f} = 
\frac{ f(p)   \Big( \frac{2}{3} f''( p) -1\Big)}{ \frac{1}{3} \big(f'( p)\big)^2 - f( p)} -1
\eeq
 The  definitions of $\rho_f$ and $P_f$ ensure that our eoms can be written 
 in the standard GR form, expressed in terms of $a$, $\rho_{\rm m}$, $\rho_{ f}$ and $P$.

Finally, we will need the standard definitions of some cosmological distances. 
$D_H$ is defined as the inverse Hubble parameter
\beq\label{def9}
D_H = \frac{1}{H} = -\frac{3}{f'( p)}
\eeq
while 
\beq\label{def10}
D_M = \int_0^z \frac{dz'}{H(z')} = \int^{t_0}_t dt' \; \frac{a(t_0)}{a(t)} =
\int^{ p}_{ p_0} \frac{d p}{ f( p_0)} \; 
\Big( \frac{f( p_0)}{f( p)}\Big)^{2/3}
\eeq
and
\beq\label{def11}
D_V = \sqrt[3]{z D_H D_M^2}.
\eeq  
represents a kind of average of the various distances and the so-called 
angular diameter is often defined as 
\beq\label{def12}
\theta = \frac{r_s}{D_V},
\eeq
where $r_s$ is the co-moving sound horizon at $t_{\rm drag}$. $\theta$ is an
important variable, e.g.\ in the observations of baryon acoustic oscillations (BAO).
Since the physics associated with these oscillations takes place before $t_{\rm LS}$ and 
thus $t_{\rm drag} < t_{\rm LS}$,  $r_s$ will  be independent of the 
functions $f$ we consider, for reasons to be discussed below.

In these formulas the coupling constants $\kp$, $\lam$ and $g$ are so far 
arbitrary, and so are $t_0$ and $t_{\rm LS}$ (or equivalently, via \rf{j42}) $ p_0$ and
$ p_{\rm LS}$). If we from observations are given $H_0$, 
the Hubble parameter at present time $t_0$,
and $z_{\rm LS}$\footnote{Of course, what is given by observations  is the temperature 
$T(t_0)$ of the CMB. $T(t_{\rm LS})$ can be calculated by atomic physics and is to a large 
extent independent of the cosmological model, as is also the statement that 
$T(t_{\rm LS})/T(t_0) = a(t_0)/a(t_{\rm LS}) = 1+z_{\rm LS}$.}, the redshift at the time of last scattering $t_{\rm LS}$, we can determine $ p_0$ and $p_{\rm LS}$:
\beq\label{j44}
f'( p_0) = - 3 \, H_0,  \quad f( p_{\rm LS}) = f( p_0) (1+z_{\rm LS})^3.
\eeq
This finally leads to the determination of $t_0 - t_{\rm LS}$ by eq.\ \rf{j42}.
We need more (experimental) input to determine the coupling constants that 
enter in $f( p)$. We will discuss this in the next section. 

Let us now consider the  three choices of $f( p)$ in \rf{j30}, where 
$f( p)$ is given by \rf{j38}. The first choice corresponds to $g=0$, $\lam > 0$:
\beq\label{j50}
\fds( p) =\frac{3}{4} \, p^2 - \lam.
\eeq
This is of course just the $f( p)$ associated with the standard de Sitter Hamiltonian 
\rf{j4}. We include this choice in order to compare the late 
time cosmology of the 
standard $\Lam$CDM model with the results from our new cosmological 
models. The second choice corresponds to $\lam =0$ and $g >0$. Thus 
\beq\label{j51}
\fg( p) = \frac{3}{4}\,(p +\al)  \sqrt{ ( p-\al)^2 +2 \al^2 }, 
\qquad \al = g^{1/3}.
\eeq
This is our model where baby universes of any size $v$ can merge with our Universe.
The third choice  is motivated by expanding $\fg(p)$ in powers of $\al/ p$:
\beq\label{j52}
\fg(p) = 
\frac{3}{4} \left( p^2 + \al^{2} \left(\frac{2\al}{ p} + 
O\Big(\frac{\al^2}{p^2} \Big)\right)\right) 
\eeq
 At early times, $v(t)$ becomes small and according to \rf{j30}
$\rho(v(t))$ will be large. Thus $\cH (v,p) =0$ implies that $| p |$ is large at early times
(i.e. times larger than but close to $t_{\rm LS}$). In addition we expect that in our Universe
$g$ is small according to \rf{j10}. Thus we expect that  keeping only the first term in
the expansion \rf{j52} is a good approximation except at quite late times where 
the exponential expansion will dominate and where $ |p|$ is close to 
$ g^{1/3}$, and this was one of the reasons we in earlier work considered 
this approximation, which when inserted in \rf{j40} led to what we denoted the 
{\it modified Friedmann equation}. We thus define this $f( p)$ as follows
\beq\label{j53}
 \fm( p) =\frac{3}{4}\,\left( p^2 + \frac{2 g}{ p}\right).
 \eeq 
 The last term in \rf{j53} is effectively a time dependent cosmological constant. At very 
 late time, when the matter term $\rho(v) \propto 1/v$ plays no role, the modified 
 Friedmann equation implies that $ p = - (2g)^{1/3}$ and the term acts
 precisely as a cosmological constant, as already shown in \rf{j9}, and in analogy with 
 the $-\lam$ term in \rf{j50}. However, 
 for smaller $t$, $ |p|$ will increase, as discussed above, and the last term in 
 \rf{j53} will be less important.
 
 The three functions $\fds$, $\fm$ and $\fm$ are so simple that even the integrals 
 appearing in \rf{j42} and \rf{def10} can be expressed as known analytic 
 functions and in this sense the models are fully solvable. In an Appendix we 
 present some details and point out a curious symmetry of the $\fm$-model.
 
Returning to the expansion \rf{j35} that defined the inclusive Hamiltonian, it is seen 
that the approximation \rf{j53} corresponds to $\phi(v) = \phi_0 + O(v)$, i.e.\ 
to the absorption of baby universes of infinitesimal size (which is one reason we 
introduced the word ``baby universe'' for the universes absorbed by our Universe). 
The absorption of such small baby universes can only result in  a change
of the spatial volume if infinitely many are absorbed per unit time 
and unit spatial volume.  

\section{Comparison of the different models}

In the three models we introduced above there are two coupling constants. The gravitational
coupling constant $\kp$ is fixed by local experiments and we will not discuss it any further.
$\fds( p)$ has the cosmological coupling constant $\lam$ as a parameter, while 
$\fm( p)$ and $\fg( p)$ have the universe-merging  coupling constant $g$ as parameter.
$\lam$ or $g$ are actually determined already from the values  $H_0$ and $z_{\rm LS}$
mentioned above provided we can assume they play no role for $t < t_{\rm LS}$. We know 
this is true for $\lam$ in the $\Lam$CDM model and from \rf{j10}, which is of course 
not exactly true when we add matter to the model, we nevertheless expect 
that it is also true for $ g$ in the other models. This implies that 
\beq\label{j61}
f( p) \approx \frac{3}{4} \, p^2 \quad {\rm for } \quad 0 < t < t_{\rm LS}.
\eeq
That being the case, we can 
calculate  $t_{\rm LS}$ as well as the relation between $t_{\rm LS}$ and $ p_{\rm LS}$ 
without reference to the specific models. For the purpose of illustration, assume 
(incorrectly) that we can ignore the radiation density $\rho_{\rm r}(t)$ all the way 
down to $t \approx 0$.  Eq.\ \rf{j42} then leads to 
\beq\label{j60}
t_{\rm LS} = \int_{-\infty}^{ p_{\rm LS}}  \frac{d  p}{\frac{3}{4} \,p^2} = 
-\frac{4}{3} \, \frac{1}{p_{\rm LS}}.
\eeq
Eqs.\ \rf{j44} determine our coupling constant since the first equation determines 
$ p_0$ and the last equation (the generalized Friedmann equation at $t_{\rm LS}$)
can be written as 
\beq\label{j62}
f( p_0) = \frac{3\, p_{\rm LS}^2}{ 4 \,(1+z_{\rm LS})^3} 
\eeq
So given $t_{\rm LS}$, and thus $ p_{\rm LS}$,  
we can determine either $\lam$ from $\fds( p)$, or $g$ from 
$\fm( p)$ or $\fg( p)$.

The $t_{\rm LS}$ used in cosmology is obtained by also including the radiation density.
In this case we obtain an additional term in eq.\ \rf{j42} from $\prt H/\prt v$ and 
\beq\label{j63}
t_{\rm LS} =  \int_{-\infty}^{ p_{LS}}   \frac{d  p}{\frac{3}{4}\, p^2 + 
\frac{1}{3} \,\kp \rho_{\rm r}(v( p))}, \qquad \frac{3}{4}\, p^2 = 
\kp \rho_{\rm m}(v) + \kp \rho_{\rm r}(v).
\eeq
This $t_{\rm LS}$ is smaller than the one given in \rf{j60} and leads to a somewhat different 
$ p_{\rm LS}$, which we should use in \rf{j62} to determine the coupling constants.
We can effectively compensate for the smaller $t_{\rm LS}$ by shifting the origin of time.

Having determined the coupling constants we can now compare the different models
for given $H_0$ and $z_{\rm LS}$. While $z_{\rm LS}$ is fixed (almost) model independent
to 1090,
there are presently two values of $H_0$ that do not agree within $5\sg$ (a fact
that is denoted the $H_0$ tension). One value
is deduced from ``local'' measurements, using various space candle techniques
\cite{h0sc}. We denote 
it $H_0^{\rm SC}$. This value ($H_0^{\rm SC} = 73.04 \pm 1.04$ km/s/Mpc) 
is almost independent of cosmological models. 
The other value is deduced from the CMB data created at 
 $t_{\rm LS}$. It is using  cosmological models in a number of ways, among 
those  to extrapolate to present time $t_0$. This value \cite{planck} 
 ($H_0^{\rm CMB} = 67.4 \pm 0.5$ km/s/Mpc), which we denote $H_0^{\rm CMB}$, is model dependent and the value usually referred to is based 
on the $\Lam$CDM model.
 
 \begin{figure}[t]
\vspace{-1cm}
\centerline{
\scalebox{1.3}{\rotatebox{0}{\includegraphics{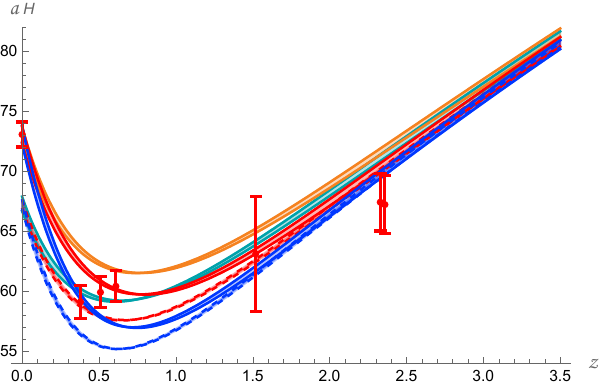}}}}
\caption[fig1]{{\small
$a(z) H(z) = H(z)/(1+z)$ shown for the three models for $H_0 = 73.0$ (case A). We have included
in red the error bars around this point, corresponding to $z=0$, but the curves all start 
at 73.0. The orange curve corresponds to $\fds$, the red curve to $\fg$ and the 
blue curve to $\fm$. In case B the three models all start at     
$H_0 = 67.4$, the value suggested by the Planck Collaboration. The green curve 
corresponds to $\fds$, the dashed red curve to $\fg$ and the dashed blue curve to $\fm$.
Inserted in red are data from $0< z < 3$:
 the first three points from the baryon acoustic oscillation data \cite{BAO}, 
the next from quasars \cite{quasars} and the last two data points from Ly--$\alpha$ measuments \cite{lyalpha1,lyalpha2}
}}
\label{fig1}
\end{figure}
 
When comparing our two models with the $\Lam$CDM model we will only use the 
CMB data to determine $T_0$, the temperature of the microwave background radiation 
at the present time. We have not attempted to use the wealth of information contained in 
the CMB temperature fluctuations since it requires the use of metric perturbations  and it is presently not clear how to incorporate such perturbations in our extended models.
We will simply present some of the variables we listed in the previous Section, namely
$H(z)$, $D_V(z)$ and $w_f(z)$, and $t_0$, calculated for the three cosmological models, in 
two cases, namely for $(H_0^{\rm SC},z_{\rm LS})$ (case A) 
and for $(H_0^{\rm CMB},z_{\rm LS})$
(case B). For $H(z)$ and $D_V(z)$ we can also 
 compare with measurements  from which one can extract  these observables 
 for relatively low values of $z$. 

In Fig.\ \ref{fig1} we have shown $a(z) H(z) = H(z)/(1+z)$ 
in case A and B for the three models.
It is seen that in case A the $\fg$ and the $\fm$ models fit the data better than 
the $\fds$ model (the $\Lam$CDM model), while in case B the opposite is the case.
Of course we do not know yet whether $H_0^{\rm SC}$ or $H_0^{\rm CMB}$ is the 
correct $H_0$ value, but {\it if} $H_0^{\rm SC}$ turns out to be the correct value the 
 $\fg$ and the $\fm$ models seems a better choice for the data presented
 here. Of course there are many other data to take into account if one wants
 to make a multi-parameter fit, but as already remarked most these data should 
 be matched to the cosmological model using perturbation expansions not yet 
 available for the $\fg$ and $\fm$ models. In Table 1 we have listed 
 the reduced $\chi^2$ values just calculated from the data and the corresponding 
 error bars shown  in Fig. \ref{fig1} in Case A and B.\\
 
 \begin{center}
 \begin{tabular}{|c|c|c|c|} \hline
 ~& $\chi_{\rm red}$(ds) & $\chi_{\rm red}$(gcdt)& $\chi_{\rm red}$(mod)\\
 \hline
 case A & 3.5 & 1.5 & 1.8\\
 \hline
 case B  & 1.2 & 1.7 & 5.6\\
 \hline
 \end{tabular}\\
 
 \vspace{6pt}
 Table 1.
 \end{center}

\begin{figure}[h]
\centerline{
\scalebox{0.55}{\rotatebox{0}{\includegraphics{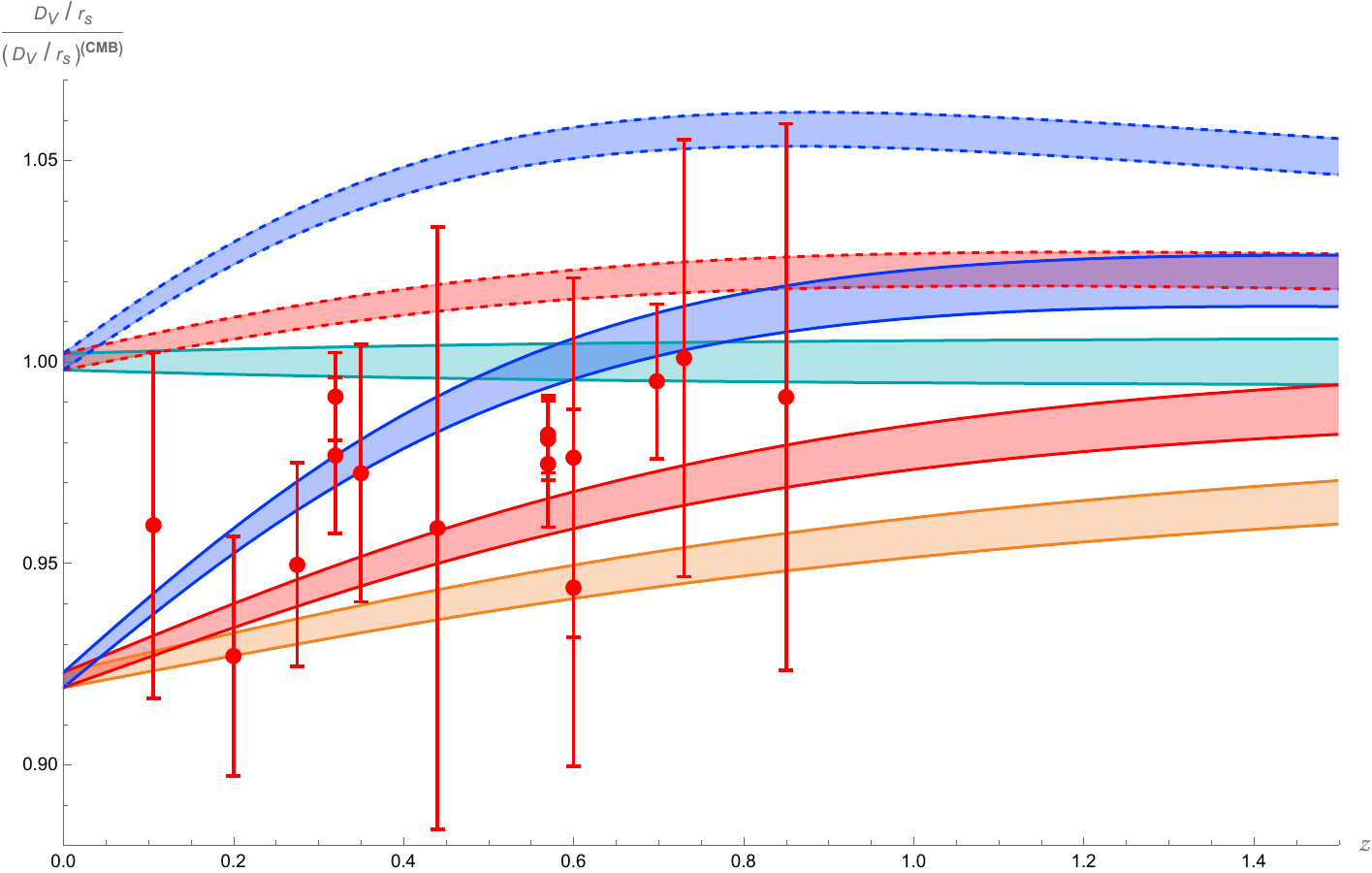}}}}
\caption[fig2]{{\small
The three curves starting at value 1 for $z=0$ are curves of 
$(D_V^{f}(z)/r_s)/(D_V^{\fds}(z)/r_s)^{\rm CMB}$ 
for the models with $f = \fg$ (red, dotted), 
$\fm$ (blue, dotted) 
and $\fds$ (green), all for the chosen value $H_0=H_0^{\rm CMB}$ (case B). 
The last curve is of course 1, except for 
the error bars reflecting the uncertainty of $H_0^{\rm CMB}$.
The three curves starting at value 0.94 for $z=0$ are the curves for $f = \fg$ (red), $\fm$ (blue) and $\fds$ (orange), all for $H_0=H_0^{\rm SC}$ (case A), and again 
normalized by $(D_V^{\fds}(z)/r_s)^{\rm CMB}$. The data points 
and error bars are obtained from  various BAO surveys (see \cite{aw3} for a 
detailed table).
}}
\label{fig2}
\end{figure}

Fig.\ \ref{fig2} shows $D_V(z)/r_s$ for the $\fds$, $\fg$ and $\fm$ models,
normalized by  $(D_V(z)/r_s)^{\rm CMB}$, i.e.\ the  $D_V(z)/r_s$
for the $\fds$ model in case B.
We see here the same trend as in Fig.\ \ref{fig1}: the $\fg$ and the $\fm$ models fit the 
data better in case A, while the $\fds$-model fits better in case B. In Table 2 
we list the reduced $\chi^2$ values obtained from Fig.\ \ref{fig2}:\\

 \begin{center}
 \begin{tabular}{|c|c|c|c|} \hline
 ~& $\chi_{\rm red}$(ds) & $\chi_{\rm red}$(gcdt)& $\chi_{\rm red}$(mod)\\
 \hline
 case A & 4.7 & 2.1 & 1.0\\
 \hline
 case B  & 1.6 & 4.1 & 13.7 \\
 \hline
 \end{tabular}\\
 \vspace{6pt}
 
 Table 2.
 
 \end{center}

\begin{figure}[t]
\centerline{
\scalebox{0.8}{\rotatebox{0}{\includegraphics{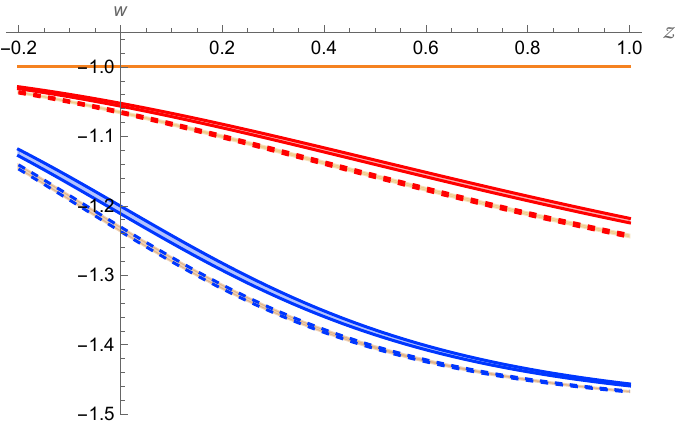}}}}
\caption[fig3]{{\small
$w_f(z)$ 
for the three models. The orange curve is $w_{\fds} =-1$, which is true  for 
any value of $H_0$. The red curves are for the $\fg$-model, dotted curve
for     $H_0^{\rm SC}$ (case A) and full curve for $H_0^{\rm CMB}$ (case B).
The blue curves are for the $\fm$-model, dotted curve
for     $H_0^{\rm SC}$ (case A) and full curve for $H_0^{\rm CMB}$ (case B).
For $z\to -1$ ($t \to \infty$) $w_f(z) \to -1$.
}}
\label{fig3}
\end{figure}

In Fig.\ \ref{fig3} we display $w_f(z)$ for the three models. Of course $w_{\fds}(z) =-1$.
The two other models have negative $w_f(z) < -1$. In standard cosmology this is a sign that
some unphysical degrees of freedom have been added to the system. However, 
here one cannot conclude that, since merging with baby universe seems more like
having a time dependent cosmological constant, but without the problem that
a time dependent  cosmological constant will break the invariance of the model 
under time-reparametrization.  Allowing a time dependence of
 the cosmological ``constant'': $\lam \to \lam(t)$, changes  $w_{\fds}$:
\beq\label{j64}
w_{\fds}(z) = -1  \to w_{\tilde{f}_{\rm ds}} = - 1 - \frac{1}{3 H(t)} \, \frac{\dot{\lam}(t)}{\lam(t)}.
\eeq
Thus, if $\lam(t)$ is growing with time, assuming the universe is expanding (i.e.\ 
$H(t) > 0$), it follows that $ w_{\tilde{f}_{\rm ds}}(t) < -1$. If we consider the 
$\fm$ model then $\lam(t)$ is replaced by $-3g/2 p$, and effectively
it acts in the same way: for small $t \to 0$ $ p(t) \to -\infty$ while for 
$t\to \infty$ $ p(t) \to - (2 g)^{1/3}$. In fact using \rf{def8} it follows 
immediately that $w_{\fm}(z)$ goes monotonically from $-3/2$ at $z=\infty$ ($t =0$),
to $-1$ for $z=-1$ ($t= \infty$), as also illustrated in Fig.\ \ref{fig3}.

Finally, in Table 3 we list the various values of $t_0$ in the six cases discussed.

\vspace{6pt}

\begin{center}
 \begin{tabular}{|c|c|c|c|} \hline
 ~& $t_0({\rm ds})$ & $t_0({\rm gcdt})$ &  $t_0({\rm mod}) $\\
 \hline
 case A & 13.3~{\rm  Gyr} & 13.5 ~ {\rm Gyr} & 13.9~ {\rm Gyr}\\
 \hline
 case B  & 13.8~ {\rm  Gyr} & 14.0~ {\rm Gyr} & 14.4~ {\rm Gyr} \\
 \hline
 \end{tabular}\\
 
 \vspace{6pt}
 
 Table 3.
 \end{center} 

 \vspace{6pt}
 
 \noindent
 In case A $t_0({\rm ds}) = 13.3\; {\rm Gyr}$  and 
 $t_0({\rm gcdt}) = 13.5\; {\rm Gyr}$ seem uncomfortable short and in case B
 $t_0({\rm mod}) = 14.4\; {\rm Gyr}$ is probably too long. However, had we started out
 with a $H_0$ in between  $H_0^{\rm SC}$ and $H_0^{\rm CMB}$ we can obtain
 values of both $t_0({\rm mod})$ and  $t_0({\rm gcdt})$ that are comfortable in agreement
 with the age of the universe as determined from the oldest stars. In particular 
 for the $\fg$-model, just looking at the data in Figs.\ \ref{fig1} and \ref{fig2}
 it is clear that if we tried to determine the best value of $H_0$ for the model from the 
 data (including now the observed $H_0^{\rm SC}$ as a data point), the optimal 
 value of $H_0$ would precisely be between  $H_0^{\rm SC}$ and $H_0^{\rm CMB}$.
In Table 4 we have listed values of $H_0$ obtained for the three model by minimizing 
the $\chi^2$ for the data shown in Fig.\ \ref{fig1}, together with the corresponding 
(minimal) value of the reduced $\chi^2$: 

\vspace{6pt}

\begin{center}
 \begin{tabular}{|c|c|c|c|} \hline
 ~& $\fds$-model  & $\fg$-model & $\fm$-model\\
 \hline
 $H_0$ & 71.2 & 72.2 & 73.9\\
 \hline
 $\chi_{\rm red}$ & 2.4  & 1.2& 1.4 \\
 \hline
 $t_0$ & 13.5 Gyr&  13.6 Gyr &  13.8 Gyr \\
 \hline
 \end{tabular}\\
 
 \vspace{6pt}
 
 Table 4.
 \end{center} 

\vspace{6pt}
 
\begin{figure}[t]
\centerline{
\scalebox{0.8}{\rotatebox{0}{\includegraphics{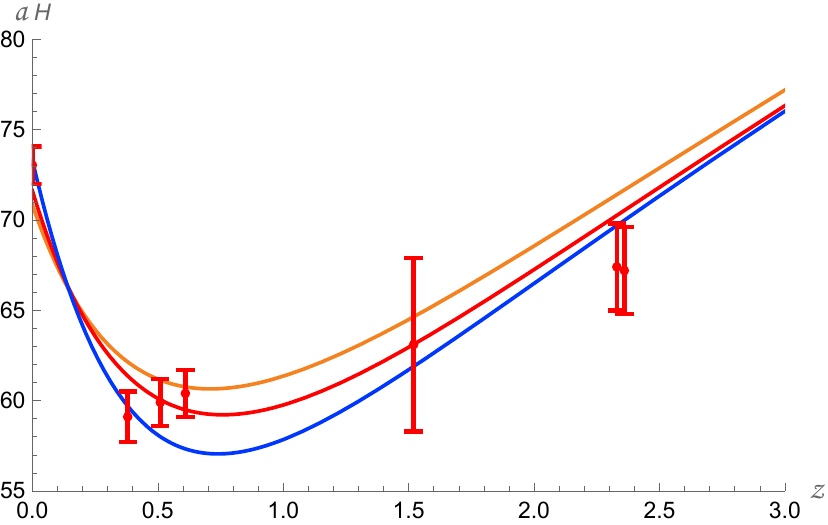}}}}
\caption[fig3]{{\small
The graphs for $H(z)/(1+z)$ for the three models using the values of $H_0$ from 
Table 4. Orange graph corresponds to $\fds$-model, blue graph to the $\fm$-model 
and red graph to the $\fg$-model.
}}
\label{fig4}
\end{figure}

 \noindent
 The $\fg$- and $\fm$-models are doing slightly better than the $\fds$-model,
 but the error bars in Fig.\ \ref{fig1} are too large to obtain any precision determination
 of $H_0$ in this way. The same is even more true if one had used the data in Fig.\ \ref{fig2}
 to determine the ``best value'' of $H_0$ for the three models. In Fig.\ \ref{fig4}
 we have shown the equivalent of Fig.\ \ref{fig2}, using the values of $H_0$ from Table 4.

We have here considered the ``locally'' observed $H(z)$ and $D_V(z)$. One could also
consider the  variable $f_m(z) \sigma_8(z)$, used for the study of density 
fluctuation of matter for $z< 2$. In \cite{aw3} this variable is defined precisely and 
calculated for the three models and compared to measured values. The error bars
in the data were too large to distinguish between the models. However, this could 
change dramatically with new observations to be obtained with the Euclid satellite. 

\section{Discussion}

Let us start by emphasizing the extreme simplicity of the models considered.
It is based on the Hamiltonian 
\beq\label{j70}
\cH = N {v} \big( -f( p) + \kp \rho_{\rm m}(v) \big),
\qquad \rho_{\rm m}(v) v = \rho_{\rm m}(v_0)v_0.
\eeq
and the corresponding eoms are
\beq\label{j71}
f(p) = \kp \rho_{\rm m}(v),\qquad \frac{\dot{v}}{v} = -f'( p), \qquad  \dot{p} = f( p).
\eeq
Thus, (for suitable $f$), if the universe starts out with a small spatial volume $v(t)$ 
for small $t$, $\rho_{\rm m}(v(t))$ is large and the first equation implies that $- p$ is 
large. The second equation then implies that the expansion is fast, and the last equation
that $- p$  increases, i.e. $|p|$ decreases, which again implies that the rate 
of expansion per unit volume $\dot{v}/v$ decreases. Eventually, if $v(t)$ continues to 
grow to infinity,  $\rho_{\rm m}(v)$ goes to zero. Thus $f( p)$ goes to zero, i.e. $ p$ 
goes to a constant. The middle equation then implies that $v(t)$ approaches an pure
exponential expansion. As we have argued, assuming the absorption of baby universes
this exponential expansion is very natural if the chance of such an absorption per 
unit time is proportional to the spatial volume.  Therefore, there is  no need to introduce a 
cosmological constant (``negative gravity'') by hand: the gravitational force indeed wants
to limit the expansion of the universe, but this is counteracted by the ``bombardment''
of our Universe by baby universes.  

The model we have suggested is of course quite primitive and unrealistic, but taking 
two limits, one where only universes with infinitesimal spatial volumes are absorbed, 
the $\fm$ model, and 
one where universes of any size can be absorbed and where our Universe 
and the other universes are on equal footing, the $\fg$ model,  
show that the models are reasonable insensitive to the detailed distribution of 
baby universe sizes. One can therefore hope that they reflect the results one would 
obtain from more realistic models.

We have presented our model as a ``late time'' cosmological model and in the present 
formulation it has nothing to say about the universe for $t < t_{\rm LS}$. Viewed as such 
a late time cosmological model it is interesting that it favors the local measured 
value $H_0^{\rm SC}$ somewhat compared to the value $H_0^{\rm CMB}$. However, in view 
of the simplicity of the model, we will not press this as an  important point. Let us rather 
discuss if this multi-universe scenario has a chance to answer some of the early-time 
questions in cosmology. 

We have nothing to add about inflation as it is presented in various models. However,
the fact that the universe has expanded  from, say,  a Planckian  size to 
$10^{-5}$m in a very short time, invites the suggestion that this expansion was 
caused by a collision with a larger universe, i.e.\ that it was really our Universe
which was absorbed in another ``parent'' universe.  Since we have presently no 
detailed description of the absorption process, it is difficult to judge if such a scenario 
could take place in a way that would actually solve the problems inflation was designed 
to solve, but one interesting aspect of such a scenario is that there is no need for an 
inflaton field.

While  a continuous  absorption of microscopic 
baby universes probably can be accommodated in a non-disruptive way in 
our Universe, it is less clear what happens if the ``baby'' universe is not small, since 
we have not suggested an actual mechanism for such absorption. Maybe the least 
disruptive situation would be one where the absorption happened inside a black
hole. The unknown mechanism of absorption could maybe favor such a scenario when
the sizes of the baby universes are not infinitesimal. 
Recall that a Reissner-Nordstr\"{o}m black hole actually 
connects to different universes. We are not seriously suggesting such a black hole 
scenario, but we mention
it to point out that there is room for a lot of interesting considerations. Ultimately,
any realistic model should be specific about how the absorption occurs.

The present value of the cosmological constant $\lam$ in the $\Lam$CDM model 
is according to some viewpoints embarrassingly  small.
 According to  \rf{j10}, $\lam$ being small when expressed
in Planck units ($\kp = 1$) also implies that $g$ is small. Historically, before 
observations pointed to a small $\lam$ in the $\Lam$CDM model,  many people 
favored $\lam=0$ as a result of some underlying mechanism not yet fully understood,
e.g.\ Coleman's mechanism \cite{coleman}. We might 
still need such a mechanism to explain why 
a $\lam$ coming  from the zero-point fluctuations of quantum fields
will not create a large $\lam$ \footnote{There are other 
viewpoints where the value of $\lam$ is claimed to be natural (or at least not 
embarrassingly small) in a quantum field context when one uses  renormalization
group arguments \cite{cosmo-constant} (see \cite{review1} for a review). }.  
However, it might be easier to explain why $\lam$
is exactly zero than to explain why it is unnaturally  small. If $\lam=0$ can be proven,
we could then still have an exponentially expanding universe caused by baby 
universe absorption. 

Like $\lam$, $g$ appears in our model just as a coupling constant, reflecting in some 
way the ``density'' of the baby universes ``surrounding'' our Universe. By 
comparing our model to observations $ g$ has to be small. However,
it is a coupling constant in a larger multiverse theory that we have not yet been 
able to solve. This leaves the hope that a consistent solution of this larger 
theory might determine $g$. We have started the program to unveil this  theory
\cite{aw1}, but it is still work in progress.

\vspace{12pt}
\section*{ Acknowledgments}

JA thanks the Perimeter Institute for hospitality while this work was completed and 
the research was supported in part by the Perimeter Institute for Theoretical Physics.
Research at the Perimeter Institute is supported by the Government of Canada throuh the Department of Innovation, Science and Economic Development and by the Province of 
Ontario through the Ministry of Colleges and Universities.

YW acknowledges the support from JSPS KAKENHI Grant Number JP18K03612.

\section*{Appendix}

In this Appendix we list some of the analytical results for the three models
defined by $\fds$, $\fg$ and $\fm$, respectively. As remarked above 
we have  already analytic parametric representations of  all the observables considered,
using $\kp p$ as parameter. Here we will provide the explicit expressions and also
provide some of them as analytic functions of time $t$.

\subsection*{The $\fds$-model}

$\fds(p)$ is given \rf{j50}. From \rf{j42} we can find $t$ as a function of $p$:
\beq\label{ap1}
t = \int_{-\infty}^{ p} \frac{d p'}{\fds( p')} = - 
\frac{2}{\sqrt{3 \lam}} \tanh^{-1} \Big(  \frac{2\sqrt{\lam}}{\sqrt{3}} \, \frac{1}{ p} \Big)
\eeq
\beq\label{ap1a}
p =  -\frac{2\sqrt{\lam}}{\sqrt{3}} \coth \Big( \frac{\sqrt{3 \lam}}{2} t\Big).
\eeq
\beq\label{ap2}
\fds( p) = \frac{\lam}{ \sinh^{2}\Big( \frac{\sqrt{3\lam}}{2} \, t\Big)}, \quad {\rm i.e}
\quad v(t) = \frac{\kp \rho_{\rm m0} v_0}{\lam} \; \sinh^2\Big( \frac{\sqrt{3\lam}}{2} \, t\Big)
\eeq
The only non-trivial function is $D_M$ defined by eq.\ \rf{def10}. The integral can be 
expressed by hypergeometric functions and one representation is 
\bea\label{ap3}
D_M &=& \frac{6}{\sqrt{3 \lam}} \left. \Big( \frac{\lam}{\kp \rho_{\rm m0}} \Big)^{1/3}
 \sinh^{1/3} \Big( \frac{\sqrt{3\lam}}{2} t' \Big) 
\!~_2F_1\Big[ \frac{1}{6},\oh;\frac{7}{6}; - \sinh^{2} \Big( \frac{\sqrt{3\lam}}{2} t' \Big)\Big]
\right |_t^{t_0}\\
&=& \frac{6}{\sqrt{3 \lam}}\, \Big( \frac{\lam}{\kp \rho_{\rm m0}} \Big)^{1/3}
 \Big( \frac{\lam}{\fds( p')}  \Big)^{1/6} 
\!\!~_2F_1\left.\Big[ \frac{1}{6},\oh;\frac{7}{6}; - \frac{\lam}{ \fds( p')}\Big]
\right |_{ p}^{ p_0} \label{ap4}
\eea
where 
\beq\label{ap5}
\frac{\lam}{\kp \rho_{\rm m0}} = \frac{\lam}{\fds (p_0)} = 
\sinh^{2}\Big( \frac{\sqrt{3\lam}}{2} \, t_0\Big).
\eeq

\subsection*{The $\fg$-model} 

$\fg( p)$ is given by \rf{j51}. 
From \rf{j42} we can find $t$ as a function of $p$:
\beq\label{ap6}
t =\int_{-\infty}^{ p} \frac{d  p'}{\fg( p')} = \frac{4}{3 \sqrt{6}\,\al} \left( \tanh^{-1}  
 \Big( \frac{ \sqrt{\frac{2}{3}} ( - p +2\al)}{\sqrt{( p-\al)^2 + 2\al^2}}\Big)
 - \tanh^{-1} \Big(  \sqrt{\frac{2}{3}}\Big)\right)
\eeq
Definining $t_c$ by 
\beq\label{ap7}
\frac{ 3 \sqrt{6} \, \al}{4}\;t_c= \tanh^{-1} \Big( \sqrt{ \frac{2}{3}}\Big) = 
\cosh^{-1} \sqrt{3} = \sinh^{-1} \sqrt{2},
\eeq
we can invert \rf{ap6} to find:
\beq\label{ap8}
 p= -\al \,\frac{ \sinh \frac{ 3 \sqrt{6} \, \al}{4} (t+t_c) 
 +2 \sqrt{2}}{ \sinh \frac{ 3 \sqrt{6} \, \al}{4} (t+t_c)  -\sqrt{2}}
\eeq
\beq\label{ap9}
\fg( p) = \frac{ 9 \sqrt{3} \,\al^2}{2} \, 
\frac{ \cosh \frac{ 3 \sqrt{6} \, \al}{4} (t+t_c)}{\big( \sinh \frac{ 3 \sqrt{6} \, \al}{4} (t+t_c) 
 -\sqrt{2}\big)^2}
 \eeq
Again the only non-trivial function is $D_M$ defined by  eq.\ \rf{def10}. 
In this case the integral in eq.\ \rf{def10} is   a generalized hypergeometric functions, a so-called Appell-$F_1$ function:
\beq\label{ap10}
D_M= \frac{1}{\fg( p_0)^{1/3}} \; \Big(\frac{ p' + \al}{6 \al^2}\Big)^{1/3} 
F_1 \left.\Big( \frac{1}{3}, \frac{1}{3}, \frac{1}{3}, \frac{4}{3}; 
x^{+}( p'), x^{-}( p')\Big)\right|_{ p}^{ p_0}
\eeq
where
\beq\label{ap11}
\fg( p_0) = \kp \rho_{\rm m0}, \qquad x^{\pm}( p)= 
\frac{2\pm i\sqrt{2}}{3 \al}\,(  p+\al).
\eeq

\subsection*{The $\fm$-model} 

$\fm(p)$ is given by \rf{j52}. 
From \rf{j42} we can find $t$ as a function of $p$:
\beq\label{ap12}
t =\int_{-\infty}^{ p} \frac{d  p'}{\fm( p')} =
\frac{4}{3 \sqrt{3}\, a}\Big( \arctan \frac{2 p -a}{\sqrt{3}\, a} +  
\frac{1}{2\sqrt{3} }\log \frac{ ( p-a)^2 +a p}{(p+a)^2}   + \frac{ \pi}{2}\Big)
\eeq
where 
\beq\label{ap13}
a= 2^{1/3} \al,  \qquad t \to 0 ~{\rm for}~ p \to -\infty, \qquad t \to \infty ~{\rm for} ~ 
 p \to -a.
\eeq
Again we can express $D_M$, defined by the integrals in eq.\ \rf{def10} ,
in terms of hypergeometric functions. One representation is:
\beq\label{ap14}
D_M = \frac{3}{5} \Big(\frac{2}{3}\Big)^{2/3} \frac{1}{ \al ^2 f(p_0)^{1/3}}
\; ( p')^{5/3} \!~_2F_1 \left.
\Big( \frac{5}{9},\frac{2}{3};\frac{14}{9}; -\frac{( p')^3}{2 \al^3}\Big)\right|_{ p}^{ p_0}.
\eeq

\subsection*{Duality in the $\fm$-model}

We end by observing that we have a kind of duality in the $\fm$-model.
Define a (modified) Hubble parameter by $h(t) = \dot{v}/v= 3 \dot{a}/a = 3 H(t)$, 
where $H(t)$ is the usual Hubble parameter. We then have the equations:
\beq\label{ap15}
\frac{3}{4} \Big( p^2 + \frac{2 \al^3}{ p} \Big) = \kp \rho_{\rm m}(v),
\quad
h (v) = -\frac{3}{4} \Big(2  p -\frac{2 \al^3}{ p^2}\Big).
\eeq
These equations are invariant under the replacement
\beq\label{ap16}
 p \;\longleftrightarrow\; -\frac{\g}{\al} \frac{\al^3}{ p}, \quad\qquad
\kp  \rho_{\rm m} \;\longleftrightarrow \;\frac{\al}{\g} \, h, \qquad \g = 2^{1/3}.
\eeq

The mapping $ p \to -\g \al^2 / p$ has some similarity with a $T$-duality map 
in string theory.


\begin{thebibliography}{99}


\bibitem{aw2}
J.~Ambj\o{}rn and Y.~Watabiki,\\
``A modified Friedmann equation,''\\
Mod. Phys. Lett. A \textbf{32} (2017) no.40, 1750224
[arXiv:1709.06497 [gr-qc]].\\
``Easing the Hubble constant tension,''\\
Mod. Phys. Lett. A \textbf{37} (2022) no.07, 2250041,
[arXiv:2111.05087 [gr-qc]].
\bibitem{aw3}
J.~Ambjorn and Y.~Watabiki,\\
``The large scale structure of the Universe from a modified Friedmann equation,''
Mod.Phys.Lett. A,  \textbf{38} (2023) no.11, 
 [arXiv:2208.02607 [gr-qc]].




\bibitem{aw1}
  J.\ Ambjorn and Y.\ Watabiki,\\
  ``A model for emergence of space and time,''\\
  Phys.\ Lett.\ B {\bf 749} (2015) 149,
  [arXiv:1505.04353 [hep-th]].\\
  ``Creating 3, 4, 6 and 10-dimensional spacetime from W3 symmetry,''\\
  Phys.\ Lett.\ B {\bf 770} (2017) 252,
  [arXiv:1703.04402 [hep-th]].\\
  ``CDT and the Big Bang,''\\
  Acta Phys.\ Polon.\ Supp.\  {\bf 10} (2017) 299,
  [arXiv:1704.02905 [hep-th]].\\
``Models of the Universe based on Jordan algebras,''\\
Nucl. Phys. B \textbf{955} (2020), 115044,
[arXiv:2003.13527 [hep-th]].

\bibitem{newkawai}
Y.~Hamada, H.~Kawai and K.~Kawana,\\
``Baby universes in 2d and 4d theories of quantum gravity,''\\
JHEP \textbf{12} (2022), 100
[arXiv:2210.05134 [hep-th]].


\bibitem{4dcdt}
J.~Ambjorn, A.~Gorlich, J.~Jurkiewicz and R.~Loll,\\
``Planckian Birth of the Quantum de Sitter Universe,''\\
Phys. Rev. Lett. \textbf{100} (2008), 091304,
[arXiv:0712.2485 [hep-th]].\\
``The Nonperturbative Quantum de Sitter Universe,''\\
Phys. Rev. D \textbf{78} (2008), 063544,
[arXiv:0807.4481 [hep-th]].

\bibitem{review}
J.~Ambjorn, A.~Goerlich, J.~Jurkiewicz and R.~Loll,\\
``Nonperturbative Quantum Gravity,''\\
Phys. Rept. \textbf{519} (2012), 127-210,
[arXiv:1203.3591 [hep-th]].\\
R.~Loll,\\
``Quantum Gravity from Causal Dynamical Triangulations: A Review,''\\
Class. Quant. Grav. \textbf{37} (2020) no.1, 013002,
[arXiv:1905.08669 [hep-th]].

\bibitem{al}
J.~Ambjorn and R.~Loll,\\
``Nonperturbative Lorentzian quantum gravity, causality and topology change,''\\
Nucl. Phys. B \textbf{536} (1998), 407-434,
[arXiv:hep-th/9805108 [hep-th]].


\bibitem{sft}
J.~Ambjorn, R.~Loll, Y.~Watabiki, W.~Westra and S.~Zohren,\\
``A String Field Theory based on Causal Dynamical Triangulations,''\\
JHEP \textbf{05} (2008), 032,
[arXiv:0802.0719 [hep-th]].

\bibitem{gcdt}
J.~Ambjorn, R.~Loll, W.~Westra and S.~Zohren,\\
``Putting a cap on causality violations in CDT,''\\
JHEP \textbf{12} (2007), 017,
[arXiv:0709.2784 [gr-qc]].

\bibitem{allgenus}
J.~Ambjorn, R.~Loll, W.~Westra and S.~Zohren,\\
`Summing over all Topologies in CDT String Field Theory,''\\
Phys. Lett. B \textbf{678} (2009), 227-232,
[arXiv:0905.2108 [hep-th]].\\
J.~Ambjorn, R.~Loll, Y.~Watabiki, W.~Westra and S.~Zohren,\\
``A New continuum limit of matrix models,''\\
Phys. Lett. B \textbf{670} (2008), 224-230,
[arXiv:0810.2408 [hep-th]].

\bibitem{inclusive}
H.~Kawai, N.~Kawamoto, T.~Mogami and Y.~Watabiki,\\
``Transfer matrix formalism for two-dimensional quantum gravity and fractal structures of space-time,''\\
Phys. Lett. B \textbf{306} (1993), 19-26,
[arXiv:hep-th/9302133 [hep-th]].

\bibitem{lectures}
J.~Ambjorn,\\
 "{Elementary Quantum Geometry}",
  arXiv:2204.00859\\
  or \\
  ``Elementary Introduction to Quantum Geometry"\\
  published 2022 by  CRC Press,
 ISBN 9781032335551



\bibitem{h0sc}
A.~G.~Riess, W.~Yuan, L.~M.~Macri, D.~Scolnic, D.~Brout, S.~Casertano, D.~O.~Jones, Y.~Murakami, L.~Breuval and T.~G.~Brink, \textit{et al.}\\
``A Comprehensive Measurement of the Local Value of the Hubble Constant with 1 km s$^{-1}$ Mpc$^{-1}$ Uncertainty from the Hubble Space Telescope and the SH0ES Team,''\\
Astrophys. J. Lett. \textbf{934} (2022) no.1, L7,
arXiv:2112.04510 [astro-ph.CO].

\bibitem{planck}
N. Aghanim et al. (Planck), \\
“Planck 2018 results. VI.
Cosmological parameters,” \\
Astron. Astrophys. 641,
A6 (2020), arXiv:1807.06209 [astro-ph.CO].

\bibitem{BAO}
Shadab Alam et al. (BOSS),\\
 “The clustering of galaxies in the completed SDSS-III Baryon Oscillation Spec- 
troscopic Survey: cosmological analysis of the DR12 galaxy sample,” \\
 Mon. Not. Roy. Astron. Soc. 470, 2617–2652 (2017), arXiv:1607.03155 [astro-ph.C
O].

\bibitem{quasars}
Pauline Zarrouk et al., \\
“The clustering of the SDSS-IV extended Baryon Oscillation Spectroscopic Survey DR14 quasar sample: measurement of the growth rate of structure from the anisotropic correlation function between redshift 0.8 and 2.2,” \\
Mon. Not. Roy. Astron. Soc. 477, 1639–1663 (2018), arXiv:1801.03062 [astro- ph.CO].

\bibitem{lyalpha1}
J.~E.~Bautista, N.~G.~Busca, J.~Guy, J.~Rich, M.~Blomqvist, H.~d.~Bourboux, M.~M.~Pieri, A.~Font-Ribera, S.~Bailey and T.~Delubac, \textit{et al.}\\
``Measurement of baryon acoustic oscillation correlations at $z=2.3$ with SDSS DR12 Ly$\alpha$-Forests,''\\
Astron. Astrophys. \textbf{603} (2017), A12,
arXiv:1702.00176 [astro-ph.CO].

\bibitem{lyalpha2}
A.~Font-Ribera \textit{et al.} [BOSS],\\
``Quasar-Lyman $\alpha$ Forest Cross-Correlation from BOSS DR11 : Baryon Acoustic Oscillations,''\\
JCAP \textbf{05} (2014), 027,
arXiv:1311.1767 [astro-ph.CO].


\bibitem{cosmo-constant}
C.~Moreno-Pulido and J.~Sola Peracaula,\\
``Renormalizing the vacuum energy in cosmological spacetime: implications for the cosmological constant problem,''\\
Eur. Phys. J. C \textbf{82} (2022) no.6, 551,
[arXiv:2201.05827 [gr-qc]].\\


C.~Moreno-Pulido and J.~Sola,\\
``Running vacuum in quantum field theory in curved spacetime: renormalizing $\rho_{vac}$ without $\sim m^4$ terms,''\\
Eur. Phys. J. C \textbf{80} (2020) no.8, 692,
[arXiv:2005.03164 [gr-qc]].


\bibitem{review1}
J.~Sola Peracaula,\\
``The cosmological constant problem and running vacuum in the expanding universe,''\\
Phil. Trans. Roy. Soc. Lond. A \textbf{380} (2022), 20210182,
arXiv:2203.13757 [gr-qc].

\bibitem{coleman}
  S.~R.~Coleman,\\
  ``Why There Is Nothing Rather Than Something: 
  A Theory of the Cosmological Constant,''\\
  Nucl.\ Phys.\ B {\bf 310} (1988) 643.
  

\end{thebibliography}
\end{document}